\title{Generalizing the Kodama State I: Construction}
\author{
        Andrew Randono \\
               Center for Relativity, Department of Physics\\
        University of Texas at Austin \\
        Austin, TX 78712\\
                  email: arandono@physics.utexas.edu
}

\documentclass[12pt]{article}
\usepackage{amssymb}
\newcommand{\beq}{\begin{equation}}
\newcommand{\eeq}{\end{equation}}
\newcommand{\beqa}{\begin{eqnarray}}
\newcommand{\eeqa}{\end{eqnarray}}
\newcommand{\w}{\wedge}
\newcommand{\g}{\gamma}
\newcommand{\ts}{\textstyle}
\newcommand{\nn}{\nonumber}

\begin{document}
\maketitle
\bibliographystyle{utphys}
\begin{abstract}
The Kodama State is unique in being an exact solution to all the ordinary constraints of canonical quantum gravity
that also has a well defined semi-classical interpretation as a quantum version of a classical spacetime, namely
(anti)de Sitter space. However, the state is riddled with difficulties which can be tracked down to the
complexification of the phase space necessary in its construction. This suggests a generalization of the state to real
values
of the Immirzi parameter. In this first part of a two paper series we show that one can generalize the state to real
variables and the result is surprising in that it appears to open up an infinite class of physical states. We show that
these
states closely parallel the ordinary momentum eigenstates of non-relativistic quantum mechanics with the Levi-Civita
curvature playing the role of the momentum. With this identification, the states inherit many of the familiar
properties of the momentum eigenstates including delta-function normalizability. In the companion paper we will
discuss the physical interpretation, CPT properties, and an interesting connection between the inner product and the
Macdowell-Mansouri formulation of general relativity. 
\end{abstract}
\section{Introduction}
Perturbative techniques in quantum field theory and their extension to quantum gravity are unparalleled in
computational efficacy. In addition, because one can always retreat to the physical picture of particles as small field
perturbations propagating on a classical background, perturbation theory maximizes the ease of transition from quantum to
classical mechanics, and many processes can be viewed as quantum analogues of familiar classical events.
However, the transparent
physical picture disappears in systems where the distinction between background and perturbation to said background
is blurred. Such systems include strongly interacting systems, such as QCD, or systems where there is no preferred
background structure, such as general relativity. In contrast, non-perturbative and background independent approaches
to quantum gravity do not
distinguish background from perturbation, and are, therefore, appropriate for modeling the quantum mechanical ground
state of the universe itself that, it is hoped, will serve as the vacuum on which perturbation theory can be based.
However, this is often at the expense of losing the smooth transition from a quantum description to its classical or
semi-classical counterpart as evidenced, for example, by the notorious problem of finding the low energy limit of Loop
Quantum Gravity. The sticking point is that pure quantum spacetime may be sufficiently divorced from our classical
understanding of fields on a smooth Riemannian manifold, that matching quantum or semi-classical states with classical
analogues may be extremely difficult.

In this respect the Kodama state is unique. Not only is the state an exact solution to all the constraints of canonical
quantum gravity, a rarity in itself, but it also has a well defined physical interpretation as the quantum analogue of a 
familiar classical spacetime, namely de Sitter
or anti-de Sitter space depending on the sign of the cosmological constant\cite{Kodama:original, Kodama:original2,
Smolin:kodamareview}. Thus, the state is a candidate for the fulfillment of one of the
distinctive advantages of a non-perturbative approach over perturbative techniques: the former has the potential to
predict the purely
quantum mechanical ground state on which perturbation theory can be based. In addition, the Kodama state has many
beautiful
mathematical properties relating the seemingly disparate fields of abstract knot theory and quantum field theory on a
space of connections\cite{Witten:knots}. In particular, the exact form of the state is known in both the connection
representation
where it is the exponent of the Chern-Simons action, and in the q-deformed spin network representation where it is a
superposition of all spin networks with amplitudes given by the Kauffman bracket\footnote{The loop transform is well
understood and rigorous at the level of mathematical physics for Euclidean signature spacetime. For Lorentz signature
spacetime, the loop transform is believed to be the Kauffman bracket, but the proof requires integrating along a real
contour in the complex plane, and it is not as rigorous as in the real case (see e.g. \cite{Pullin:Book}). The de Sitter
state that we will
present shares loop transform properties in common with the Euclidean signature Kodama state, so it is well defined in
the loop basis.}\cite{Kauffman:TemperleyLieb}.
This connection played a pivotal
role in the development of the loop approach to quantum gravity. One offshoot of the connection between the state and
knot theory is that the relation with quantum
groups allows for a reinterpretation of the role cosmological constant as the modulator of the deformation parameter
of the quantum deformed group. 

Ultimately, however, observation and experiment are the arbiters of the relevance of a physical theory, and
cosmological evidence suggests that we live in an increasingly vacuum dominated universe, which is asymptotically
approaching
de Sitter space in the future as matter fields are diluted by the expansion of the universe, and possibly in the past as
well as evidenced by the success of inflation models. Thus, the state with positive $\lambda$ is particularly relevant to
modern
cosmology, and it opens up the possibility of making uniquely quantum mechanical predictions of a cosmological nature.

\subsection{Problems}
Despite all of these positive attributes of the Kodama state, the state is plagued with problems. Among these are the
following:
\begin{itemize}
\item \textbf{Non-normalizability:} The Kodama state is not normalizable under the kinematical inner product, where
one simply integrates $|\Psi|^{2}$ over all values of the complex Ashtekar connection. The state is not known to be
normalizable under a physical inner product defined by, for example, path integral methods. Linearized perturbations
around the state are known to be non-normalizable under a linearized inner product\cite{Smolin:linearkodama}.
\item \textbf{CPT Violation:} The states are not invariant under CPT\cite{Witten:note}. This is particularly poignant
objection in view
of the CPT theorem of perturbative quantum field theory, which connects CPT violation with Lorentz violation. It is not
known if the result carries over to non-perturbative quantum field theory, but it has yet to be demonstrated that the
Kodama state does not predict Lorentz violation.
\item \textbf{Negative Energies:} It has been argued by analogy with a similar non-perturbative Chern-Simons state of
Yang-Mills theory that the Kodama state necessarily contains negative energy sectors\cite{Witten:note}. If the energy of
one sector of the state is strictly positive, the CPT inverted state will necessarily contain negative energy sectors.
\item \textbf{Non-Invariance Under Large Gauge Transformations:} Although the state is invariant under the small gauge
transformations generated by the quantum constraints, it is not invariant under large gauge
transformations where it changes by a factor related to the winding number of the map from the manifold to the gauge
group. However, it has been argued that the non-invariance of the Kodama state under large gauge transformations give
rise to the thermal properties of de Sitter spacetime\footnote{Paradoxically, we will argue the opposite: that demanding
invariance of the generalized states we will present under large gauge transformations gives rise to evidence of
cosmological horizons, which in turn should give rise to the thermal nature of de Sitter space.}\cite{Soo:thermalkodama}.
Thus, non-invariance under large gauge transformations could be a problem or a benefit, but it is deserving of mention.
\item \textbf{Reality Constraints:} The Lorentzian Kodama state is a solution to the quantum constraints in the Ashtekar
formalism where the connection is complex. To obtain classical general relativity one must implement reality
conditions which ensure that the metric is real. It is an open problem as to how to implement these constraints on a
general state. Generally it is believed that the physical inner product will implement the reality constraints,
but this could change the interpretation of the state considerably.
\end{itemize}
\subsection{Resolution} 
Many of the above problems can be tracked down to the
complexification of the phase space necessary in the construction of the state. To see this, one can simply appeal to the
Euclidean
version of the state. In the Euclidean formalism, the gauge group $SO(4)$ splits into two left and right pieces as in
the complex theory. Choosing the left handed part of the group, the canonical variables in the Ashtekar formalism consist
of a \textit{real} $SO(3)$ connection and its \textit{real} conjugate momentum. The analogous state in the Euclidean
theory is a pure phase since the connection is real:
\begin{equation}
\Psi[A]=\mathcal{N}e^{-i\frac{3}{4k\lambda}\int Y_{CS}[A]}.
\end{equation}
Although the state may not be strictly normalizable, one might expect that it is delta-function normalizable because
it is pure phase. In fact, it has been shown that linearized perturbations to the Euclidean state are delta-function
normalizable under a linearized inner product\cite{Smolin:linearkodama}. In addition, the state is CPT invariant due to
the factor of
$i$ in the argument which inverts under time reversal canceling the action of parity. Although it is not known if the
state has negative energies, one cannot use the standard argument that a positive energy sector will become a negative
energy
sector under CPT reversal, because the action of CPT is now trivial. Since the state is now pure phase, the level of the
Chern-Simons theory is real. Thus, by fine tuning Newton's constant or the cosmological constant (within observational
error), one can make the level an integer, in which case the state is invariant under large gauge transformations.
Finally, there are no reality conditions in the Euclidean theory since the connection and its conjugate momentum are
real. Thus, the Euclidean state appears to be free of most of the known problems associated with the Lorentzian state.
However, the real world is Lorentzian: \textit{can one salvage the Lorentzian Kodama state despite all these
problems?} 

The above properties of the Euclidean state suggest that the problems associated with the Lorentzian Kodama state
are rooted in the complexification of the phase space. The phase space is complex because of a particular choice for a
free parameter, the Immirzi parameter $\beta$, which is chosen to be the unit imaginary, $-i$, in the complex
Ashtekar formalism. Modern formulations of Loop Quantum Gravity assume that $\beta$ is an arbitrary \textit{real}
number\cite{Barbero}. The parameter is currently believed to be fixed by demanding consistency with the spin network
derivation of the entropy of an isolated
horizon, and Hawking's formula for the entropy of a static, spherically symmetric black hole\cite{Ashtekar:entropy}. The
first few sections of this paper will be devoted to generalizing the state to real values
of the Immirzi parameter. The discussion will initially follow along the the lines of \cite{Randono:GK}, and then will
diverge, addressing some deficiencies of that initial attempt at generalizing the Kodama state. We
will show that generalizing the state opens up a large Hilbert space of states each
parameterized by a particular configuration of the three-dimensional Riemannian curvature. By exploiting an analogy
between these states and the ordinary momentum eigenstates of single particle quantum mechanics we will show that the
states are delta function normalizable and orthogonal unless they are parameterized by the same 3-curvature modulo
$SU(2)$ gauge and diffeomorphism transformations. Using this property we will show that the states can be used to
construct a natural Levi-Civita curvature operator. When this operator is used in the Hamiltonian constraint, all
of the states are annihilated by the constraint.
In a follow-up paper we will then show that the generalized
states are free of most of the problems associated with the
original incarnation of the Kodama state, and we will discuss the physical interpretation of the new states and their
relation to de Sitter space. We conclude with an intriguing relation between the physical inner product of two
generalized states and the Macdowell-Mansouri formulation of gravity.

\section{Chiral Asymmetric Extension of the Kodama State }
\subsection{Chirally Asymmetric Gravity}
Following along the lines of \cite{Randono:GK}, we begin the construction of the states using a chirally
asymmetric, complex action. This will allow us to make headway in generalizing the state to arbitrary \textit{imaginary}
values of the Immirzi parameter. Later we will analytically extend the states to real values of the Immirzi parameter.
The starting point for the construction of the generalized Kodama states is the Holst action with a cosmological
constant\footnote{Throughout the paper we will work with a Lorentzian metric with signature $\eta=diag(-1,1,1,1)$. The
metric volume form $\epsilon_{IJKL}$ is defined such that $\epsilon_{0123}=-\epsilon^{0123}=+1$. Upper case Roman
indices $\{I,J,K,L,...\}$ represent spacetime indices in the adjoint $Spin(3,1)$ representation space and range from $0$
to $3$. Lower case Roman indices $\{i,j,k,...\}$ are three dimensional indices in the adjoint representation of $SU(2)$, and
range from $1$ to $3$. In the base manifold, spacetime indices are represented by Greek letters
$\{\mu,\nu,\alpha,\beta,...\}$, and spatial indices are represented by lower case Roman indices in the beginning of the
alphabet $\{a,b,c,...\}$.}\cite{Holst}:
\beq
S_{H} = \frac{1}{k}\int_{M}\star e\w e\w R +\frac{1}{\beta}e\w e\w R-\frac{\lambda}{3}\star e\w e\w e\w e. 
\eeq
Here $e=\frac{1}{2}\gamma_{I}e^{I}$ is the frame field, $R=\frac{1}{4}\gamma_{[I}\gamma_{J]}R^{IJ}$ is the curvature
of the $Spin(3,1)$ connection $\omega=\frac{1}{4}\gamma_{[I}\gamma_{J]}\omega^{IJ}$, $k=8\pi G$, and 
$\star=-i\g_{5}=\g^{0}\g^{1}\g^{2}\g^{3}$. The parameter $\beta$ is the Immirzi parameter, which can be interpreted as
the measure of parity violation built into the framework of quantum gravity. The parameter does not affect the
equations of motion in the absence of matter, and matter fields can be appropriately modified so that the resulting
equations of motion reproduce that of the Einstein-Cartan action action\cite{Randono:Torsion}. The parameter does play
a significant role in the quantum theory where it fine tunes the scale where Planck scale discreteness occurs. At this stage we will
take the parameter to be purely imaginary and later analytically extend to real values. The reason we begin
with imaginary $\beta$ is because in this case the Holst action splits into two independent left and right handed components.
In this sense, imaginary values of $\beta$ can not only be interpreted as a measure of parity violation, but more
specifically they measure the degree of \textit{chiral} asymmetry built into the framework of gravity. To see this, we
introduce the left and right handed chiral projection operators $P_{L/R}=\frac{1}{2}(1\mp i\star)$, and define the
chirally asymmetric Einstein-Cartan action (writing $\Sigma \equiv e\w e$):
\beqa
S &=& \frac{1}{k}\int_{M}2(\alpha_{L} P_{L}+\alpha_{R}P_{R})
\star \Sigma \w \left(R-\ts{\frac{\lambda}{6}}\Sigma\right) \nn \\
&=& \frac{2}{k}\int_{M}\alpha_{L}\star\Sigma_{L}\w \left(R_{L}-\ts{\frac{\lambda}{6}}\Sigma_{L}\right)
+ \alpha_{R}\star\Sigma_{R}\w \left(R_{R}-\ts{\frac{\lambda}{6}}\Sigma_{R}\right) \nn\\
&=& \frac{1}{k}\int_{M}(\alpha_{L}+\alpha_{R})\star \Sigma \w \left(R-\ts{\frac{\lambda}{6}}\Sigma\right)
 + i(\alpha_{L}-\alpha_{R})\Sigma \w R.
\eeqa
The last line is the Holst action if we make the identifications $\alpha_{L}+\alpha_{R}=1$ and 
\beq
\beta=\frac{-i}{\alpha_{L}-\alpha_{R}}.
\eeq
We note that in the limiting case when $\alpha_{R}=0$ and $\beta=-i$ we recover the left handed Einstein-Cartan action
whose phase space consists of the complex left-handed Ashtekar action and its conjugate momentum. The advantage of the
this formalism is that the action splits into two components that, prior to the implementation of reality constraints,
can be treated independently. The reality constraint requires that $e^{I}$ and $\omega^{IJ}$ are real. This implies the
constraints $\Sigma^{IJ}_{L}=\overline{\Sigma^{IJ}_{R}}$ and $\omega^{IJ}_{L}=\overline{\omega^{IJ}_{R}}$. We will
proceed to construct the quantum constraints and a generalization of the Kodama state initially assuming all left
handed variables are independent of right handed variables. Later we will impose the above reality
constraints.

Proceeding to construct the constraints assuming left and right handed variables are independent we find that the
constraint algebra spits into two independent copies which differ by handedness and by the relative coupling constants
$\alpha_{L}$ and $\alpha_{R}$. We demand that the manifold has topology $\mathbb{R}\times \Sigma$ where 
$\Sigma$ is the spatial topology. Introducing a monotonic time function $t$, we define a timelike vector field
$\bar{t}$ to be the canonical dual of the one form $dt$, so that $dt(\bar{t})=1$. We further split the vector field
into components normal and parallel to the 3-manifold which we denote $\bar{t}=N\bar{n}+\bar{N}$. The vector field
$\bar{n}$ is the unit normal to $\Sigma$, $N$ is the lapse, and $\bar{N}$ is the shift. We will partially fix the gauge to
the time gauge where $e^{0}_{a}=0$ with $a=\{1,2,3\}$ components in the base manifold. This is achieved by fixing the
direction of the unit normal in the fibre so that $e^{I}(\bar{n})=n^{I}=(1,0,0,0)$. This is not strictly
necessary in the Ashtekar formalism since self dual $spin(3,1)$ variables in $M$ can be pulled back to complex $su(2)$
valued variables in $\Sigma$. However, we will work in the time gauge in order to make contact with the real Ashtekar-Barbero
formalism where gauge fixing is necessary. The left and right handed connections pullback naturally to $\Sigma$ to
form the canonical position variable $A^{ij}_{L}=\omega^{ij}+i K^{ij}$ and 
$A^{ij}_{R}=\omega^{ij}-i K^{ij}$. Here $K^{ij}={\epsilon^{ij}}_{k}K^{k}$ and $K^{i}$ is the extrinsic curvature
defined by $\phi^{*}Dn^{I}$ where $\phi^{*}$ is the pullback of the map from $\Sigma$ to $M$. In our gauge, the
extrinsic curvature is then $K^{i}_{a}={\omega^{i}}_{0a}$. The canonical momenta to $\omega^{ij}_{L}$ and
$\omega^{ij}_{R}$ are the two forms $-\frac{i\alpha_{L}}{2k}\Sigma^{L}_{ij}$ and 
$\frac{i\alpha_{R}}{2k}\Sigma^{R}_{ij}$. Eventually we will want 
$\Sigma^{ij}_{L}=\Sigma^{ij}_{R}=E^{i}\w E^{j}$ where $E^{i}_{a}\equiv e^{i}_{a}$ is the spatial triad in the time gauge,
but for now we are treating the two as independent. With this, the canonical commutation relations are
\beqa
\left\{A^{ij}_{L}|_{P}, \Sigma^{L}_{kl}|_{Q}\right\}
&=&-\frac{2k}{\alpha_{L}}\ \delta^{[i}_{m}\delta^{j]}_{n}\ \delta(P,Q)\nn \\
\left\{A^{ij}_{R}|_{P}, \Sigma^{R}_{kl}|_{Q}\right\}
&=& \frac{2k}{\alpha_{R}}\ \delta^{[i}_{m}\delta^{j]}_{n}\ \delta(P,Q)\nn\\
\left\{A^{ij}_{L}|_{P}, \Sigma^{R}_{kl}|_{Q}\right\}&=&\left\{A^{ij}_{R}|_{P}, \Sigma^{L}_{kl}|_{Q}\right\}=0
\label{CCR1}
\eeqa

Each of the constraints contain two independent left and right handed components:
\beqa
C_{H}(N)&=&\alpha_{L}\int_{\Sigma}N \left(*\Sigma^{L}_{ij}\w\left(R^{ij}_{L}-{\ts
\frac{\lambda}{3}}\Sigma^{ij}_{L}\right)\right)
\ +\ (L\rightarrow R) \\
C_{G}(\lambda_{L},\lambda_{R})&=&\alpha_{L}\int_{\Sigma}D_{L}\lambda^{L}_{ij}\w \Sigma^{ij}_{L}\ -\ (L\rightarrow R)
\\
C_{D}(\bar{N})&=&\alpha_{L}\int_{\Sigma} \mathcal{L}_{\bar{N}}A^{ij}_{L}\w \Sigma^{L}_{ij}\ -\ (L\rightarrow R)
\eeqa
The Hamiltonian constraint, $C_{H}$, generates time reparameterizations through the shift, $N$, the ``Gauss" or ``gauge"
constraint,
$C_{G}$, generates infinitesimal $SU_{L}(2)\times SU_{R}(2)$ transformations with $\lambda_{L}$ and $\lambda_{R}$ as
generators, and
the diffeomorphism constraint, $C_{D}$, generates infinitesimal three-dimensional diffeomorphisms along the vector field
$\bar{N}$.
We now need to promote the constraints to quantum operators. We will work in the connection representation where the
momenta are functional derivatives: 
\beqa
\Sigma^{L}_{ij}=\frac{2k}{\alpha_{L}}\frac{\delta}{\delta \omega^{ij}_{L}} 
& & \Sigma^{R}_{ij}=-\frac{2k}{\alpha_{R}}\frac{\delta}{\delta \omega^{ij}_{R}}.
\eeqa
Since the left and right handed variables are independent the Hilbert space also splits into two copies: $\mathcal{H}_{R}\times
\mathcal{H}_{L}$. Thus we will look for solutions of this form. With the operator ordering given above, the
constraints immediately admit the Kodama-like solution:
\beq
\Psi[A_{L}, A_{R}]=\mathcal{N}\exp\left[-\frac{3}{4k\lambda}\left(\alpha_{L}\int_{\Sigma}Y_{CS}[A_{L}]
-\alpha_{R}\int_{\Sigma}Y_{CS}[A_{R}]\right)\right]. \label{LRKodama}
\eeq
where the $Y[A]=A\wedge dA +\frac{2}{3}A\w A\w A $ is the Chern-Simons three-form and the
implied trace is in the adjoint representation of $su(2)$. Here we have used the fundamental identity 
\beq
\frac{\delta}{\delta A_{ij}}\int_{\Sigma}{A^{p}}_{q}\w d{A^{q}}_{p}+
\frac{2}{3}{A^{p}}_{q}\w{A^{q}}_{r}\w{A^{r}}_{p}=-2F^{ij}.
\eeq
We note that in the limit that $\alpha_{L}=1$ and $\alpha_{R}=0$, we regain the original form of the Kodama state. 

\subsection{Imposing the Reality Constraints}
We now need to impose the reality constraints $\Sigma_{L}=\overline{\Sigma_{R}}$ and 
$A_{L}=\overline{A_{R}}$. Imposing the constraints on the position variables is easy since these
are just multiplicative operators. We define the real and imaginary parts of $A_{L}$ by\footnote{Having fixed our
index conventions in the previous sections, in the remaining sections we will drop all indices. Unless stated
otherwise, we will work in the adjoint representation of $SU(2)$.}
\beqa
\omega^{ij}\equiv Re(A_{L})&=&\frac{1}{2}(A_{L}+A_{R})\\
K\equiv Im(A_{L}) &=& \frac{1}{2i}(A_{L}-A_{R}).
\eeqa
It then follows that $A_{L}=Re(A_{L})+iIm(A_{L})=\omega+iK$ and 
$A_{R}=\overline{A_{L}}=\omega-iK$. The constraint on the momentum variables is slightly more subtle due to the
partial gauge fixing we have employed. Without gauge fixing we would have $\Sigma^{ij}_{L}=e^{i}\w e^{j}
+i{\epsilon^{ij}}_{k}e^{i}\w e^{0}$, but in the time gauge $e^{0}_{a}=0$ so $\Sigma^{ij}_{L}=E^{i}\w E^{j}$ is real. To
implement this in the quantum theory we define
\beqa
\Sigma \equiv Re(\Sigma_{L}) &=& \frac{1}{2}(\Sigma_{L}+\Sigma_{R}) \\
C_{\Sigma}\equiv Im(\Sigma_{L})&=& \frac{1}{2i}(\Sigma_{L}-\Sigma_{R})=0.
\eeqa

We now need to add the constraint $C_{\Sigma}$ into the full set of constraints. We encounter a problem when
evaluating the full set of commutators---the constraint algebra no longer closes. In particular, we find that the
commutator between the Hamiltonian constraint and $C_{\Sigma}$ yields a second class constraint proportional to the
torsion of $\omega$:
\beq
\{C_{H}, C_{\Sigma}\} \sim D_{\omega}*\Sigma=T.
\eeq
Typically this second class constraint is solved at the classical level by replacing the unconstrained $SU(2)$ spin
connection $\omega$ with the torsion-free Levi-Civita connection, $\Gamma=\Gamma[E]$, where $\Gamma$ is a solution to the
torsion
condition $dE^{i}=-{\Gamma^{i}}_{k}\w E^{k}$. In our context this implies that the left and right spin connections are
replace by $A_{L}=\Gamma+iK$ and $A_{R}=\Gamma-iK$. With these replacements, the left and right handed connections will
no longer commute: $\{\omega_{L}, \omega_{R}\}\neq 0$. This is our first indication that something will go wrong with
this initial attempt at generalizing the Kodama state when the full set of constraints is employed. We will see that
we can avoid this issue entirely by a proper reinterpretation of the problem.

However, there is another, potentially more serious problem associated with the introduction of the constraint
$C_{\Sigma}$. In particular, the generalized state we have constructed does not satisfy the quantum constraint
$C_{\Sigma}\Psi = 0$. To illustrate the problem it is useful to redefine the basis of our phase space such that
$\Sigma=\frac{1}{2} (\Sigma_{L}+\Sigma_{R})$ and $C_{\Sigma}$ are the new canonical momenta up to numerical
coefficients. 
The associated canonical position variables are 
\beqa
A_{-\frac{1}{\beta}}&\equiv&\alpha_{L}A_{L}+\alpha_{R}A_{R}=\Gamma+{\ts\frac{1}{\beta}}K \\
A_{\beta}&\equiv&\frac{\alpha_{L}A_{L}-\alpha_{R}A_{R}}{\alpha_{L}-\alpha_{R}}=\Gamma-\beta K, 
\eeqa
which can be seen from the canonical commutation relation that follow directly from \ref{CCR1},
\beqa
\{A_{-\frac{1}{\beta}}, C_{\Sigma}\}&=& i2k\ \delta(P,Q) \nn\\
\{A_{\beta}, \Sigma \}&=& -i2k\beta\ \delta(P,Q) \nn\\
\{A_{-\frac{1}{\beta}}, \Sigma\}&=& 0 \nn\\
\{A_{\beta}, C_{\Sigma}\} &=& 0 \label{CCR2}.
\eeqa
We recognize $A_{\beta}$ and $\Sigma=E\w E$ as the Ashtekar-Barbero connection and its momentum that emerge in the real
formulation of LQG. The reason for introducing these variables is that the constraint $C_{\Sigma}\Psi=0$ takes a
particularly simple form. In the connection representation, $C_{\Sigma}=2k \frac{\delta}{\delta A_{-1/\beta}}$.
Thus, we must have\footnote{The limiting case when $\beta \rightarrow \mp i$ must be treated separately here because
in those cases we have an initial primary constraint that $\Sigma_{R/L}=0$.} 
\beqa
C_{\Sigma}\Psi=2k \frac{\delta}{\delta A_{-1/\beta}}\Psi =0 \ &\longrightarrow& \ \Psi=\Psi[A_{\beta}].
\eeqa
That is, the wavefunction can only be a function of the Ashtekar-Barbero connection $A_{\beta}$ and is independent of
$A_{-1/\beta}$.

Now we need to check that the state (\ref{LRKodama}) is only a function of $A_{\beta}$. To do so, we express the
state in terms of the $A_{\beta}$ and $A_{-1/\beta}$. We rewrite the state in a form that will be convenient for
later use:
\begin{equation}
\Psi[A]=\mathcal{N}\exp\left[\frac{-3i}{4k\Lambda \beta^{3}}\int_{\Sigma}Y_{CS}[A]-(1+\beta^{2})Y_{CS}[\Gamma] 
+2\beta(1+\beta^{2})Tr(K\wedge R_{\Gamma})\right]. \label{GK1}
\end{equation}
Here $\Gamma$ and $K$ are explicit functions of both $A_{\beta}$ and $A_{-1/\beta}$, given by
\beqa
\Gamma &=& \frac{A_{\beta}+\beta^{2}A_{-1/\beta}}{1+\beta^{2}}\\
K &=& \frac{1}{\beta}(\Gamma-A_{\beta}).
\eeqa
We see that the state is explicitly a function of \textit{both} $A_{\beta}$ and $A_{-1/\beta}$. Thus,
$C_{\Sigma}\Psi\neq 0$.
\subsection{Resolution}
The problems we have encountered with this initial attempt at generalizing the Kodama state are twofold. First, we
encounter a second-class constraint whose solution requires that we introduce the torsion-free spin connection
$\Gamma=\Gamma[E]$. This means that the left and right handed variables will no longer commute, or in the new
variables, $A_{\beta}$ and $A_{-1/\beta}$ will no longer commute. Second, we find that the reality constraint on the
momentum requires that the wave function is a functional of $A_{\beta}$, which is not true for our left-right
asymmetric state. We can recast the problem in a slightly more intuitive way by eliminating $A_{-1/\beta}$ in favor of
the momentum $\Sigma$. That is, we explicitly write $A_{-1/\beta}=\frac{1}{\beta^{2}}((1+\beta^{2})\Gamma-A_{\beta})$
and treat $\Gamma[E]$ as an explicit function of the momentum conjugate to $A_{\beta}$. Then the problem can be
restated, \textit{why is the wavefunction an explicit function of both position and momentum variables?} The problem
of defining the commutator of $A_{\beta}$ and $A_{-1/\beta}$ is transmuted into the problem of defining the operator
$\Gamma[E]$ which occurs explicitly in the Hamiltonian through the Levi-Civita curvature, $R_{\Gamma}$, or the extrinsic
curvature, $\frac{1}{\beta}(\Gamma-A)$, depending on how one writes the constraints. We will see that we can address
both of these problems by analytically extending the state to real values of the Immirzi parameter, $\beta$. This will
allow us to exploit an analogy between the generalized Kodama state and the non-relativistic momentum eigenstates, which
will suggest a reinterpretation of the explicit momentum dependence of the state and at the same time suggest a
natural definition of the Levi-Civita curvature operator $R_{\Gamma}$. This will be the subject of the rest of the
paper.

\section{The Generalized Kodama States}
\subsection{Properties of the real state}
We now consider the state (\ref{GK1}) when the Immirzi parameter $\beta$ is taken to be a non-zero, but otherwise
arbitrary real number. Modern formulations of Loop Quantum Gravity begin with arbitrary real values of $\beta$ in the
canonical construction because the analysis of real $SU(2)$ connections is better understood than that for complex
connections. In addition, it is believed that thermodynamic arguments will eventually fix the value of the Immirzi
parameter unambiguously. For our purposes, taking $\beta$ to be real changes the properties of the generalized Kodama state considerably.

We first address the issue of the explicit momentum dependence of the state. We appeal to a similar situation in
ordinary single particle quantum mechanics. The generalized Kodama for real values of $\beta$ shares many
properties in common with the ordinary momentum eigenstates. First
of all, both states are pure phase. This means that they are bounded, which has implications for the inner product.
Whereas the complex Kodama state is unbounded, which implies that the state is non-normalizable under a naive inner
product, the real state is pure phase and therefore may be normalizable in the strict sense if the phase space is
compact, or delta-function normalizable if the phase space is non-compact. Secondly, the momentum eigenstates share
the property in common with the generalized Kodama state in that they ostensibly depend explicitly on both the momentum
and position variables. Of course, the role of the momentum in the momentum eigenstates is very different from the
role of the position variables. The state $\Psi_{p}(x)=\mathcal{N}e^{i p\cdot x-iEt}$ is
explicitly a function of the position variable only, but it is \textit{parameterized} by the momentum $p$.
That is, the momentum eigenstates form a large family of orthogonal states distinguished by a particular value of
$p$. This is the interpretation we will adopt for the role of the momentum in the generalized Kodama state.
To see this explicitly, we rewrite the state (\ref{GK1}) in a more suggestive form by absorbing irrelevant factors which
depend only on the momentum through $\Gamma[E]$ into the normalization constant. The state becomes:
\begin{equation}
\Psi_{R}[A]=\mathcal{P}\exp\left[i\kappa\int_{\Sigma} A\wedge R-\frac{1}{2(1+\beta^{2})}Y_{CS}[A] \right].
\end{equation}
Here we see explicitly, $A$ plays the role of the position variable $x$, the Levi-Civita curvature
$R=d\Gamma +\Gamma\w\Gamma$ plays the role of the momentum, $\kappa=\frac{3(1+\beta^{2})}{2k\lambda\beta^{3}}$ is simply a scaling
factor, we have a dimensionless energy $\frac{1}{2(1+\beta^{2})}$, and the
Chern-Simons term $\int Y_{CS}[A]$ plays the role of the time variable. We note that it has been independently suggested
that the Chern-Simons invariant is a natural time variable on the canonical phase space\cite{Soo:thermalkodama}. With
this interpretation, the
generalized state is not a single state at all, but a large class of states parameterized by a specific configuration of
the three-dimensional Levi-Civita curvature, $R$. 

\subsection{The naive inner product}
We can push the analogy further by considering the inner product between two states with different curvature
configurations $\langle\Psi_{R'}|\Psi_{R}\rangle$. The analogue of this is the inner product of two momentum states:
\beqa
\langle p'|p\rangle&=&\int d^{n}x \ \Psi^{*}_{p'}[x,t]
\Psi_{p}[x,t] \nn\\
&=& \mathcal{P}[p', p]\int d^{n}x \ \exp[-i(p'-p)\cdot x]\nn\\
&\sim& \delta^{n}(p'-p).
\eeqa
Following along these lines, we define a naive inner product:
\beqa
 \langle\Psi_{R'}|\Psi_{R}\rangle_{naive} &=& \mathcal{P}[\Gamma',\Gamma]\int_{\Sigma}\mathcal{D}A\ 
  \Psi^{*}_{R'}[A]\Psi_{R}[A]\nn\\
 &=& \int \mathcal{D}A\ \exp\left[-i\kappa \int_{\Sigma} A\w (R'-R)\right] 
 \eeqa
 Formally integrating over the space of connections we have
 \beq
 \langle \Psi_{R'}|\Psi_{R}\rangle_{naive} \sim \delta(R'-R).
 \eeq
 Thus, under this naive inner product, two states are orthogonal unless they are parameterized by the same configuration
of the Levi-Civita curvature. The deficiency of this inner product is that it is not gauge invariant. If the two
fields $R'$ and $R$ represent the same curvature written in a different gauge, either $SU(2)$ or diffeomorphism, they
will be orthogonal. Thus, we need to modify the inner product to make it gauge invariant.

\subsection{Gauge covariance and the kinematical inner product}
In order to define a gauge invariant inner product we first need to discuss the gauge properties of the generalized
states. The set of states ${\Psi_{R}}$ are not strictly speaking $SU(2)$ gauge or diffeomorphism invariant.
The reason is because of the presence of the parameter $R$ in the argument which acts like an effective ``background"
against which one can measure the effect of a gauge transformation or diffeomorphism. This is not unfamiliar. We
encounter the same difficulty with the spin-network states where the graph serves as a ``background" against which one
can measure the effect of a diffeomorphism shifting the connection, $A$ (see e.g. \cite{Rovelli:book}). In the spin
network states, the action of a
one-parameter diffeomorphism $\phi_{-\bar{N}}$ on the connection configuration, $A$, is equivalent to shifting the graph
in the opposite direction
by $\phi_{\bar{N}}$. Similarly, one can show that the combined effect of an SU(2) gauge transformation and diffeomorphism
on the field configuration which we will denote by $\phi_{\{g^{-1},-\bar{N}\}}A$ is equivalent to the inverse
transformation
on the curvature denoted by $\phi_{\{g, \bar{N}\}}R$. Thus, under the action of the Gauss and diffeomorphism constraint,
the state transforms as follows:
\begin{equation}
\Psi_{R} \rightarrow  
\hat{U}_{\phi}(g^{-1},-\bar{N}) \Psi_{R} =
 \Psi_{\phi_{\{g,\bar{N}\}}R}.
\end{equation}

The strategy with the spin network states is to implement the diffeomorphism symmetry via the inner product
where the diffeomorphism symmetry is manageable, and this is the strategy we will also adopt. To make the inner product
gauge invariant, we introduce the measure
$\mathcal{D}\phi_{\{g,\bar{N}\}}$ over the set of all SU(2) gauge transformations (which may be accomplished by the
Haar measure) and the set of all diffeomorphisms. Although a measure over the set of all diffeomorphisms is undefined,
the end result may still be manageable due to the specific form of the integrand. This is true in the
inner product on spin network states, where the problem of defining a measure over the group of diffeomorphisms is
relegated to the problem of determining when two graphs are in the same equivalence class of knots. A similar result
applies here. To see this, we define the kinematical inner product as follows:
\beq
\langle\Psi_{R'}|\Psi_{R}\rangle_{kin} = 
\int \mathcal{D}\phi_{\{g,\bar{N}\}}\langle \Psi_{R'}|U_{\phi}(g,\bar{N})\Psi_{R}\rangle_{naive}\ .
\eeq
From the gauge covariance of the states $\Psi_{R}$ we have:
\beqa
\langle\Psi_{R'}|\Psi_{R}\rangle_{kin} &=& 
\int \mathcal{D}\phi_{\{g,\bar{N}\}}\langle\hat{\phi}_{\{g^{-1},-\bar{N}\}}\Psi_{R'}|\Psi_{R}\rangle \nn\\
&=& \int \mathcal{D}\phi_{\{g,\bar{N}\}}\langle\Psi_{\phi_{\{g,\bar{N}\}}R'}|\Psi_{R}\rangle \nn\\
&\sim& \int \mathcal{D}\phi_{\{g,\bar{N}\}}\delta(\phi_{\{g,\bar{N}\}}R'-R)\nn\\
&=& \delta(\mathcal{R'}-\mathcal{R})
\eeqa
where in the last line $\mathcal{R'}$ and $\mathcal{R}$ are elements of the equivalence class of curvatures modulo
$SU(2)$-gauge and diffeomorphism transformations. Thus, the problem of defining a measure over the set of diffeomorphisms
is reduced to the problem of determining when two curvatures are gauge related---a problem that is all too familiar from
classical General Relativity. The states $\Psi_{R}$ and
$\Psi_{R'}$ are orthogonal unless there is a diffeomorphism and/or SU(2) gauge transformation relating $R$ and $R'$.

\subsection{Levi-Civita curvature operator}
Continuing the analogy with the momentum eigenstates we proceed to define a Levi-Civita curvature operator. We recall
the momentum operator can be defined in terms of the momentum eigenstates:
\beq
\hat{p}=\int d^{n}p'\ p'|p'\rangle\langle p'|.
\eeq
By construction, the states $|p\rangle$ are then eigenstates of $\hat{p}$.

Since the generalized states $|\Psi_{R}\rangle$ represent a family of orthogonal states parameterized by the curvature
configuration $R$, it is natural to define a curvature operator such that the states are curvature eigenstates. Analogous to
the momentum operator, we define the operator in its diagonal form as follows (writing $\phi=\phi_{\{g,\bar{N}\}}$):
\begin{equation}
\int_{\Sigma}\alpha\wedge\hat{R}_{\Gamma}=\int \mathcal{D}\phi\mathcal{D}\Gamma '
\left[\left(\int_{\Sigma}\lambda\wedge\phi R'_{\Gamma '}\right) |\Psi_{\phi R'}\rangle
\langle\Psi_{\phi R'}|\right]
\end{equation}
where $\alpha$ is an arbitrary $su(2)$ valued one-form serving as a test function, and $\mathcal{D}\Gamma '$ is an
appropriate measure to integrate over all values of the Levi-Civita 3-curvature
$R'_{\Gamma '}$. When operating on a state $|\Psi_{R}\rangle$ it is understood that intermediate inner product is the 
naive inner product. That is,
\beqa
& &\int_{\Sigma}\alpha \wedge \hat{R}\ |\Psi_{R}\rangle \nn\\
& &=\int \mathcal{D}\phi \mathcal{D}\Gamma '
\left[\left(\int_{\Sigma}\alpha\wedge\phi R'_{\Gamma '}\right) |\Psi_{\phi R'}\rangle
\langle\Psi_{\phi R'}|\Psi_{R}\rangle_{naive}\right]\nn\\
&& =\int \mathcal{D}\phi \mathcal{D}\Gamma '
\left[\delta(\phi R'-R)\ \left(\int_{\Sigma}\alpha\wedge\phi R'_{\Gamma '}\right)
|\Psi_{\phi R'}\rangle \right] \nn\\
&& =\int_{\Sigma}\alpha\wedge R_{\Gamma}\ 
|\Psi_{R}\rangle
\eeqa

Thus, with this definition, the states
$|\Psi_{R}\rangle$ are eigenstates of the curvature operator $\hat{R_{\Gamma}}$:
\beq
\int_{\Sigma}\alpha\wedge \hat{R}\ |\Psi_{R}\rangle=\int_{\Sigma}\alpha\wedge R \ |\Psi_{R}\rangle. \label{Roperator}
\eeq

\subsection{The Hamiltonian constraint}
We now address the issue of the Hamiltonian constraint. The beauty of the complex Ashtekar formalism is that the
Hamiltonian constraint simplifies to the point where it is solvable, admitting the Kodama state as a quantum solution to
the Hamiltonian constraint. Our partial parity violating version of the Ashtekar action held the promise of a
simplified Hamiltonian until the reality constraints were imposed, which introduced second class constraints on the
torsion. When solved, the constraint implies that the left and right handed connections no longer commute because
they both
contain a term $\Gamma[E]$. The real formulation of the Holst action, is plagued with the same problem. Although the
phase space consists of just the connection $A$ and its conjugate momentum, the Hamiltonian constraint explicitly
contains terms involving $\Gamma[E]$. Depending on how one writes the constraint, they enter via extrinsic curvature
terms, $K=\frac{1}{\beta}(\Gamma-A)$, or through the Levi-Civita curvature, $R=d\Gamma+\Gamma\w\Gamma$. The standard
representation of the Hamiltonian constraint is\footnote{The term involving $*\Sigma\w D_{\Gamma}K$ may be unfamiliar
since it is usually not included in the constraint. The term does explicitly occur in the Hamiltonian decomposition,
but it can be integrated away using the fact that $\Gamma$ is torsion free so $D_{\Gamma}*\Sigma=0$. We will keep the
term explicitly because it simplifies the algebra in the next step.}
\beq
C_{H}=\int_{\Sigma}*\Sigma\w\left(F+(1+\beta^{2})({\ts \frac{1}{\beta^{2}}}D_{\Gamma}K
-K\w K)-{\ts \frac{\lambda}{3}}\Sigma \right),
\eeq
where $F=F[A]$ is the curvature of $A$. Because of the complexity of this constraint, it appears to be very difficult
to determine if our generalized Kodama
states are in the kernel of the corresponding quantum operator. However, the constraint can be rewritten by
substituting the extrinsic curvature terms in favor of the Levi-Civita curvature. The constraint then takes the form
\beq
C_{H}=\int_{\Sigma} *\Sigma\w\left((1+{\ts \frac{1}{\beta^{2}}})R-{\ts\frac{1}{\beta^{2}}}F
-{\ts \frac{\lambda}{3}}\Sigma \right).
\eeq
This form is particularly convenient for our purposes because we have already suggested a form for the Levi-Civita
curvature operator. In the standard Kodama operator ordering where $*\Sigma$ is placed on the far left, the full set
of generalized Kodama states are in the kernel of the Hamiltonian by virtue of being in the kernel of the quantum
operator
\beq
\int_{\Sigma} \alpha\w \left((1+{\ts \frac{1}{\beta^{2}}})\hat{R}-{\ts\frac{1}{\beta^{2}}}\hat{F}
-{\ts \frac{\lambda}{3}}\hat{\Sigma} \right)
 \eeq
where $\alpha$ is a test function. To see this, in the connection representation $\Sigma$ is a differential operator
which acts on $\Psi_{R}[A]$ by:
\beq
-{\ts\frac{\lambda}{3}}\Sigma \ \Psi_{R}[A]=i2 k\beta {\ts\frac{\lambda}{3}}\ \frac{\delta \Psi_{R}[A]}{\delta A} 
=\left({\ts \frac{1}{\beta^{2}}}F-(1+{\ts \frac{1}{\beta^{2}}})R\right)\ \Psi_{R}[A]\ .
\eeq
The curvature $F$ cancels since $\hat{F}$ is multiplicative in the connection representation.
We are left with 
\beq
(1+{\ts \frac{1}{\beta^{2}}}) \int_{\Sigma}\alpha\w (\hat{R}-R)\ \Psi_{R}[A]\ ,
\eeq
which vanishes by (\ref{Roperator}). Thus, for any curvature configuration, $R$, with the standard Kodama operator
ordering we have
\beq
\hat{C}_{H}\ |\Psi_{R}\rangle =0 .
\eeq
\section{Concluding Remarks}
We have shown that the Kodama state can be generalized to real values of the Immirzi parameter, and the generalization
appears to open up a large class of physical states.
In the connection
representation they are the exponent of the Chern-Simons invariant together with an extra term, so we might expect that
in the spin network basis they may be expressed as a generalization of the Kauffman bracket. The states share many 
properties in common with the momentum eigenstates when the Levi-Civita is identified with the ``momentum"
parameterizing the family of states. Following this analogy, we have shown that the generalized Kodama states are
eigenstates of a naturally defined Levi-Civita curvature operator with eigenvalues given by the curvature
configuration parameterizing the state. This definition of the curvature operator places the full sector in the physical
Hilbert space. Naturally this operator must withstand a program of consistency
checks to verify its viability as the Levi-Civita curvature operator, but we hope that we have laid the groundwork to
begin such a program. We set out to generalize the Kodama state in an attempt to resolve some of the known issues
associated with the original version. We have shown that our generalization solves two of the known problems: reality
conditions, and normalizability. The problem of defining the reality conditions does not exist in the real theory, and
we have shown that the states are delta-function normalizable under a natural inner-product. In the second paper of
this two paper series we will show that the states are $CPT$ invariant, and, by fine tuning of the
coupling constants, they can be made to be invariant under large transformations. In addition, we will discuss the
physical interpretation of the states. 

\section*{Acknowledgments}
I would like to thank Lee Smolin especially for many stimulating discussions and for his support on both papers in this
series. I would also like to thank Laurent Freidel for his criticism of my first paper which inspired this work.
\bibliography{GKSI}
\end{document}